\documentclass[runningheads, envcountsame, a4paper]{llncs}
\usepackage[hyphens]{url}
\usepackage{amsmath,amssymb}
\usepackage{graphicx}
\usepackage{tikz}
\usepackage{colortbl}
\usepackage{microtype}
\usepackage{xcolor}
\usetikzlibrary{positioning,decorations.pathreplacing,shapes}
\usepackage{subcaption}
\usepackage{soul}
\usepackage{color}
\usepackage{arydshln}
\usepackage{booktabs}
\usepackage{url}
\usepackage{verbatim}

\newcommand{\yT}{\widetilde{\mathbf{y}}}

\newcommand{\yH}{\widehat{\mathbf{y}}}
\newcommand{\bT}{\widetilde{\mathbf{b}}}
\newcommand{\bH}{\widehat{\mathbf{b}}}
\newcommand{\uH}{\widehat{\mathbf{u}}}

\newcommand{\mIcal}{\mid \mathcal{I}}
\newcommand\Rmb[1][]{\widehat{\boldsymbol{\Sigma}}_{U#1}}
\newcommand\Rb[1][]{\widehat{\boldsymbol{\Sigma}}_{B#1}}
\newcommand{\Smb}{\mathbf{A}}

\newcommand{\R}{\mathbb{R}}
\newcommand{\E}{\mathbb{E}}
\newcommand{\var}{\operatorname{Var}}

\begin{document}
	
\title{Probabilistic Reconciliation of Hierarchical Forecast via Bayes' Rule}
\author{Giorgio Corani \and Dario Azzimonti \and Joao P. S. C. Augusto \and Marco Zaffalon}
\authorrunning{Corani et al.}   
\tocauthor{Giorgio Corani (Idsia),  Dario Azzimonti (Idsia), Joao P. S. C. Augusto (Idsia), 
	Marco Zaffalon (Idsia)}
\institute{Istituto Dalle Molle di Studi sull'Intelligenza Artificiale (IDSIA)\\
	USI - SUPSI\\
	Manno, Switzerland\\
	\email{ giorgio\{dario.azzimonti,zaffalon\}@idsia.ch}
}
\maketitle  

\begin{abstract}
		We present a novel approach for reconciling  hierarchical forecasts,  based on Bayes' rule.
		We define a prior distribution for the bottom time series of the
		hierarchy, based on the bottom base forecasts. Then we
		update their distribution via Bayes’ rule,  based on the base forecasts for the
		upper time series. Under the Gaussian assumption, we  derive 
		the updating in closed-form. We derive two algorithms, which differ 
		as for the assumed independencies.
		We discuss their relation with the MinT reconciliation algorithm and with the Kalman filter, and we compare them experimentally. 
\end{abstract}

\section{Introduction}
Often time series are organized into a hierarchy.
For example, the total visitors of a country can be divided into regions and the visitors of each region can be further divided into  
sub-regions.
The most disaggregated time series of the hierarchy are referred to as 
\textit{bottom time series}, while the remaining time series are referred to as  \textit{upper time series}.

Forecasts of hierarchical time series should be \textit{coherent}; for instance, the sum of the forecasts of the different regions
should equal the forecast for the total. 
The forecasts are \textit{incoherent} if they do not satisfy such constraints.
A simple way for generating coherent forecasts is \textit{bottom-up}:  one takes the forecasts for the bottom time series and sums them up 
according to the summing constraints in order to produce forecasts for the entire hierarchy.
Yet this approach does not consider the forecasts produced for the upper time series, which contain useful
information.
For instance, upper time series are smoother  and allow to better estimate of the seasonal patterns 
and the effect of the covariates.

Thus, modern reconciliation methods \cite{Hyndman.etal2011,Wickramasuriya.etal2018}  proceed in two steps.
First, 
\textit{base forecasts} 
are computed by fitting an independent model
to each time series.
Then, the base forecasts are adjusted to become coherent; this step is called \textit{reconciliation}.
The forecasts for the entire hierarchy are then obtained by summing up the reconciled bottom time series.
Reconciled forecasts
are generally more accurate than the base forecasts, as they
benefit from information coming from multiple time series.  
The state-of-the art reconciliation algorithm is MinT \cite{Wickramasuriya.etal2018},
which minimizes the mean squared error of the reconciled forecasts by solving a 
generalized least squares problem; 
its point forecast, besides being coherent,  are generally more accurate than the base forecast.

Hierarchical probabilistic forecasting is however still an open area of research.
The algorithm by \cite{taieb17a} constructs a coherent forecast in a bottom-up fashion, modelling via copulas the joint distribution of the bottom time series, while 
\cite{2014variational} proposes a top-down approach, where the top time series is forecasted and then disaggregated.
Both algorithms are based on numerical procedures which have no closed-form solution; hence they are not easily interpretable.
In \cite{gamakumara2018probabilisitic}, a geometric interpretation of the reconciliation process is provided. It is moreover 
shown  that the log score is not proper 
with respect to incoherent probabilistic forecasts.  
As an alternative, the energy score can be used for
comparing reconciled to unreconciled probabilistic hierarchical forecasts. 
In \cite{ath2019} multivariate Gaussian predictive densities and bootstrap densities are experimentally compared for hierarchical probabilistic forecasting. 

We address  probabilistic reconciliation using Bayes' rule.
We define the prior beliefs about the bottom time series,
based on the base forecasts for the bottom time series.
We then update them incorporating the information contained 
in the forecasts for the upper time series.
Under the Gaussian assumption, we compute the update in closed form, obtaining the posterior distribution about the  bottom time series and then about the entire hierarchy.
Our reconciled forecasts minimize the mean squared error; indeed, we prove that they match the point predictions of MinT, whose optimality has been proven in a frequentist way.  
Our algorithm provides the joint predictive distribution for the hierarchy;  thus we call it pMinT, which stands for probabilistic MinT. We also provide a variant of pMinT, obtained by making an additional independence assumption; 
we call it LG, as it is related to the  linear-Gaussian model \cite[Chap.8.1.4]{bishop2006}.

We show a link between the
reconciliation problem and the Kalman filter, opening the possibility of  borrowing from the  literature of the Kalman filter for future research.
We then compare the algorithms on synthetic and real data sets, eventually concluding that 
pMinT yields more accurate probabilistic forecasts than both bottom-up and LG.

The paper is organized as follows: we introduce the reconciliation problem in 
Section~\ref{sec:reconciliationIntro}, we discuss the algorithms
in Sec. \ref{sec:GaussianReconciliation},
we discuss the reconciliation of a simple hierarchy in
Section~\ref{sec:BasicExample} and we present
the experiments in Section~\ref{sec:experiments}.  

\section{Time series reconciliation}
\label{sec:reconciliationIntro}
\begin{figure*}[!ht]
	\centering
	\begin{tikzpicture}[level/.style={sibling distance=40mm/#1}, , scale=0.9]
\node [circle,draw] (z){Total}
child {node [circle,draw] (a) {\textcolor{white}{r}$\,R_1\,$\textcolor{white}{r}}
	child {node [circle,draw, fill=gray!20] (b) {$\,R_{11}\,$}
	}
	child {node [circle,draw, fill=gray!20] (g) {$\,{R_{21}}\,$}
	}
}
child {node [circle,draw] (j) {\textcolor{white}{r}$\,R_2\,$\textcolor{white}{r}}
	child {node [circle,draw, fill=gray!20] (k) {$\,R_{21}\,$}
	}
	child {node [circle,draw, fill=gray!20] (l) {$\,R_{22}\,$}
	}
};
\end{tikzpicture}
	\caption{A hierarchical time series which disaggregates the visitors into regions and sub-regions.}
	\label{fig:hierExample}
\end{figure*}
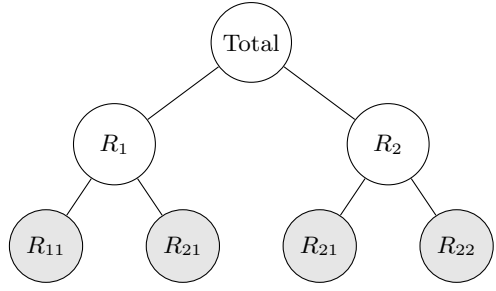

Fig.~\ref{fig:hierExample} shows a hierarchy. We could interpret it as the visitors of a country, which are disaggregated first by region (R$_1$, R$_2$) and then by sub-regions (R$_{11}$, R$_{12}$, R$_{21}$, R$_{22}$). 
 The most disaggregated time series
(\textit{bottom time series}) are shaded.
The hierarchy contains $m$ time series, of which $n$ are bottom time series.
We denote  by uppercase letters  the random variables 
and by lowercase letters their observations.
The vector of observations available at time $t$ for the entire hierarchy is $\mathbf{y}_t \in \R^m$; they are observations from the set of random variables
$\mathbf{Y}_t$.
Vector $\mathbf{y}_t$
can be broken down in
two parts, namely  $\mathbf{y}_t = [\mathbf{u}_t^T, \mathbf{b}_t^T]^T$;
$\mathbf{b}_t \in \R^n$ contains the observations of the bottom time series while
$\mathbf{u}_t \in \R^{m-n}$ contains the observations of the upper time series.
At time $t$, the observations available for the hierarchy of Fig.~\ref{fig:hierExample} are thus:
\begin{align*}
& \mathbf{y}_t = [y_{\text{Total}},y_{R_1},y_{R_2},
y_{R_{11}},y_{R_{12}},y_{R_{21}},y_{R_{22}}
]^T  = [\mathbf{u}_t^T ,  \mathbf{b}_t^T]^T,\\
&\text{where:}\\
& \mathbf{u}_t = [y_{\text{Total}},y_{R_1},y_{R_2}]^T \\
& \mathbf{b}_t = [y_{R_{11}},y_{R_{12}},y_{R_{21}},y_{R_{22}}]^T. \\
\end{align*}

The structure of the hierarchy is represented by the summing matrix $\mathbf{S} \in \R^{m\times n}$  such that:
\begin{equation}
\mathbf{y}_t = \mathbf{S} \mathbf{b}_t .
\label{eq:Reconciliation}
\end{equation}
The $\mathbf{S}$ matrix of hierarchy in Fig.\ref{fig:hierExample} is:
\newcommand\bigA{\makebox(0,0){\text{\huge{A}}}}
\newcommand\bigI{\makebox(0,0){\text{\huge{I}}}}
\begin{equation}
\mathbf{S} = \begin{bmatrix}
	1 & \,\,\, & 1 & \,\,\, & 1 & \,\,\, & 1\\
	1 & \,\,\, & 1 & \,\,\, & 0 & \,\,\, & 0 \\
	0 & \,\,\, & 0 & \,\,\, & 1 & \,\,\, & 1 \\
	1 & \,\,\, & 0 & \,\,\, & 0 & \,\,\, & 0 \\
	0 & \,\,\, & 1 & \,\,\, & 0 & \,\,\, & 0 \\
	0 & \,\,\, & 0 & \,\,\, & 1 & \,\,\, & 0 \\
	0 & \,\,\, & 0 & \,\,\, & 0 & \,\,\, & 1 \\
\end{bmatrix}
=
\begin{bmatrix}
	 & \,\,\, &  & \,\,\,   &        & \,\,\, &  \\
	 & \,\,\, &  & \bigA   &        & \,\,\, &  \\
	 & \,\,\, &  & \,\,\, & \,\,\, &        &  \\  \cdashline{2-6} \
	 & \,\,\, &  & \,\,\,   &        & \,\,\, &  \\
	 & \,\,\, &  & \bigI   &        & \,\,\, &  \\
	 & \,\,\, &  & \,\,\,   &        & \,\,\, &  \\
	 & \,\,\, &  & \,\,\,   &        & \,\,\, &
\end{bmatrix},
\label{eq:matrixS}
\end{equation}
\noindent
where the sub-matrix $\mathbf{A} \in \mathbb{R}^{(m-n)\times n}$ encodes which bottom time series should be summed up in order to obtain each upper time series.

We denote by $\yH_{t+h} \in \R^m$ the base forecasts issued at time $t$ about of $y$ and referring to $h$ steps ahead. 
We separate base forecasts for bottom time series ($\bH_{t+h} \in \R^n$) and upper time series ($\uH_{t+h} \in \R^{m-n}$), namely
$\yH_{t+h} = [\uH_{t+h}^T, \bH_{t+h}^T]^T$.  The variances of the error of the base forecasts will be used later. 
If forecasts for different time horizons are needed (e.g., $h$=1,2,3,..), 
the reconciliation is performed independently for each $h$.
In the following we generically assume to reconcile the forecasts for $h$ steps ahead.

\paragraph*{The MinT reconciliation}
Most reconciliation algorithms \cite{Hyndman.etal2011}, including
MinT \cite{Wickramasuriya.etal2018}, assume the 
reconciled bottom forecasts ($\bT_{t+h}$) to be a linear combination of the base forecasts ($\yH_{t+h}$) available for the whole hierarchy, i.e. their objective is to find a matrix  $\mathbf{P}_h \in \R^{n \times m}$ such that:
\begin{equation}
\bT_{t+h} = \mathbf{P}_h\yH_{t+h}.
\label{eq:uTilde}
\end{equation}

Let us denote by $\widehat{\mathbf{E}}_{t+h} = \mathbf{Y}_{t+h} - \yH_{t+h} \in \R^m$ the vector of the errors of the base forecast $h$-steps ahead
and by $\mathbf{W}_h = \E[\widehat{\mathbf{E}}_{t+h}\ \widehat{\mathbf{E}}^T_{t+h} \mIcal_t]$ their covariance matrix, where $\mathcal{I}_t$ denotes all the information available up to time $t$.
In \cite{Wickramasuriya.etal2018} it is proven that the reconciliation matrix given
by:
\begin{equation}
\mathbf{P}_h = (\mathbf{S}^T \mathbf{W}_h^{-1} \mathbf{S})^{-1} \mathbf{S}^T \mathbf{W}_h^{-1}
\label{eq:PMinTEstimator}
\end{equation}
is optimal, in the sense that it minimizes the trace of the reconciliation errors' covariance matrix. 
The reconciled forecasts for the whole hierarchy are obtained by summing the reconciled bottom forecasts, and they are proven  to minimize the mean squared error over the entire hierarchy.

\subsubsection{Estimation of $\mathbf{W}_h$}\label{sec:reconc-h}
Estimating $\mathbf{W}_h$
differently for each  $h$ is an open problem.
For the case $h$=1 the estimation is simpler.
The variance of the forecasts equals the variance of the residuals (i.e., the errors on 1-step predictions made on the training data) and cross-covariances are estimated as the covariance of the residuals.
The best estimates are obtained \cite{Wickramasuriya.etal2018}
by shrinking the full covariance matrix towards a diagonal matrix,
using the method of \cite{schafer2005shrinkage}.

The case $h$ \textgreater 1 is instead problematic. 
The variance of the forecasts are obtained by 
increasing the 1-step variance through analytical formulas, which differ from the 
variance of the $h$-steps ahead residuals.
Morevoer, the covariances in $\mathbf{W}_h$ have to be numerically estimated by looking at the $h$-steps residuals.
However, the number of 
$h$-steps residuals decreases with $h$, making the estimate more noisy.

As a workaround, \cite{Wickramasuriya.etal2018} assumes 
$\mathbf{W}_h = k_h \mathbf{W}_1$, where $k_h > 0$
is an unknown constant which depends on $h$ while $\mathbf{W}_1$ is the covariance matrix of the one-step ahead errors.  
The underlying assumption is thus that all terms within the variance/covariance matrix of the errors grow in the same way with $h$. The advantage of this approach is that $k_h$
cancels out when computing the reconciled forecasts,
as it can be seen by setting $\mathbf{W}_h = k_h \mathbf{W}_1$ in Eq.\eqref{eq:PMinTEstimator}.
Yet, $k_h$  appears in the expression of the variance of the reconciled forecasts. 
In the following we refer to the assumption $\mathbf{W}_h = k_h \mathbf{W}_1$ as ``the $k_h$ assumption''.

\section{Probabilistic Reconciliation}\label{sec:GaussianReconciliation}

We address the reconciliation problem 
by merging the probabilistic information contained in the 
base forecasts for the bottom and the upper time series.
We perform the fusion using Bayes' rule.

We first 
define the prior about 
the \textit{bottom} time series.
We have observed the time series up to time $t$ and we are interested in the reconciled forecasts for time $t+h$.
We denote by $\mathbf{B}_{t+h}$ the vector of the bottom time series at time $t+h$; this is thus a vector of random variables and
$B^i_{t+h}$ represents its $i$th element.
We moreover denote by $\bH_{t+h}$ the vector of base
forecasts 
for the bottom time series for  time $t+h$,
and by $\mathbf{b}_{t+h}$ the actual observation of the bottom random variables at time $t+h$.
Finally, 
$\mathcal{I}_{t,b}$ is the information available up to time $t$ regarding the bottom time series, i.e. the past values of the bottom time series: $\mathcal{I}_{t,b} =\{\mathbf{b}_1, \ldots, \mathbf{b}_t \}$.

In the following we adopt the $k_h$ assumption for all the  covariance matrices assuming moreover that, for a given $h$, the value of $k_h$  is shared among all the involved covariance matrices. As we will show later, this is equivalent to the $k_h$ assumption made by MinT.
Let us hence denote the covariance matrix of the forecast $h$-steps ahead by
$\Rb[,h] = k_h  \Rb[,1]$.
Assuming the bottom time series to be jointly Gaussian we have:
\begin{align}
p(\mathbf{B}_{t+h} \mIcal_{t,b}) &= N\left(\bH_{t+h},  k_h  \Rb[,1] \right).
\end{align}

\paragraph{Probabilistic bottom-up}
If we have no information about the upper time series, we can build a joint predictive distribution for the entire hierarchy by summing the bottom forecast via matrix $\mathbf{S}$: 
\begin{align}
p(\mathbf{Y}_{t+h} \mIcal_{t,b}) &= N\left(\mathbf{S}\bH_{t+h},  \mathbf{S}  k_h \Rb[,1] \mathbf{S}^T \right), 
\end{align}
which is a \textit{probabilistic bottom-up} reconciliation.
Note that, in this case, $k_h$ appears only in the expression of the variance.

\paragraph{Updating}
If  the forecasts $\widehat{\mathbf{U}}_{t+h}$ about the upper time series are available, then we can use them in order to
update our prior. We assume:
\begin{align}
& \widehat{\mathbf{U}}_{t+h} = \Smb \mathbf{B}_{t+h} + \boldsymbol{\varepsilon}_{t+h}^u, \label{eq:upperNoise} \\
& \boldsymbol{\varepsilon}_{t+h}^u \sim N\left(\mathbf{0}, \Rmb[,h] \right), \nonumber 
\end{align}
where $\Rmb[,h] = k_h \Rmb[,1]$ is the covariance  of the noise.
We thus treat $\widehat{\mathbf{U}}_{t+h}$
as a set of different sums of the future values of the bottom time series, corrupted by noise. Hence:
\begin{align}
p(\widehat{\mathbf{U}}_{t+h}  \mid \mathbf{B}_{t+h}) &= N\left( \Smb \mathbf{B}_{t+h}, k_h\Rmb[,1] \right). \label{eq:U=ab+e}
\end{align}

The posterior 
distribution of the bottom time series is given by Bayes' rule:
\begin{align}
p(\mathbf{B}_{t+h} \mIcal_{t,b}, \widehat{\mathbf{U}}_{t+h} ) & = 
\frac{p(\mathbf{B}_{t+h} \mIcal_{t,b}) 
	p(\widehat{\mathbf{U}}_{t+h}  \mIcal_{t,b}, \mathbf{B}_{t+h})
}{p(\widehat{\mathbf{U}}_{t+h} \mIcal_{t,b})} = 
\nonumber
\\
& = 
\frac{p(\mathbf{B}_{t+h} \mIcal_{t,b}) 
	p(\widehat{\mathbf{U}}_{t+h} | \mathbf{B}_{t+h})
}{p(\widehat{\mathbf{U}}_{t+h} \mIcal_{t,b})} 
\propto \nonumber \\
&\propto p(\mathbf{B}_{t+h} \mIcal_{t,b}) p(\widehat{\mathbf{U}}_{t+h} \mid \mathbf{B}_{t+h}) = \nonumber \\
&
=p(\mathbf{B}_{t+h} \mIcal_{t,b}) p(\Smb \mathbf{B}_{t+h} + \boldsymbol{\varepsilon}_{t+h}^u \mid \mathbf{B}_{t+h}) \label{eq:Bayes_rule}
\end{align}

\subsection{Computing  Bayes' rule}
The posterior of Eq.~\eqref{eq:Bayes_rule} can be computed in closed form by
assuming the vector $(\mathbf{B}_{t+h}, \widehat{\mathbf{U}}_{t+h})$ to be jointly Gaussian distributed. 
The linear-Gaussian (LG) model \cite[Chap.8.1.4]{bishop2006}.
computes analytically the updating
by further assuming $\boldsymbol{\varepsilon}_{t+h}^u$ to be independent from 
$\mathbf{B}_{t+h}$. 
Yet this independence might not always hold in our case. Consider for instance a special event driving
upwards most time series. 
As a result we would observe 
both high values of $\mathbf{B}_{t+h}$ and negative values of 
$\boldsymbol{\varepsilon}_{t+h}^u$, 
due to the underestimation  of the upper time series.
This would result in a correlation between 
$\mathbf{B}_{t+h}$ and 
$\boldsymbol{\varepsilon}_{t+h}^u$.

We thus  generalize  the LG model by accounting for such correlation.
We will later compare experimentally the results obtained adopting the LG model and its generalized version.
We denote 
$ \operatorname{Cov(\mathbf{B}_{t+1} ,\boldsymbol{\varepsilon}_{t+1}^u \mIcal_{t,b}})=
\mathbf{M}_{1} \in \R^{n\times(m-n)}$
and we assume
$ \operatorname{Cov(\mathbf{B}_{t+h} ,\boldsymbol{\varepsilon}_{t+h}^u \mIcal_{t,b}})=
k_h\mathbf{M}_{1}$.

Our first step for computing Bayes' rule
is to express the joint  distribution
$p(\mathbf{B}_{t+h}, \widehat{\mathbf{U}}_{t+h} \mIcal_{t,b})$. Since
$\widehat{\mathbf{U}}_{t+h} = \Smb \mathbf{B}_{t+h} + \boldsymbol{\varepsilon}_{t+h}^u$,
the expected values are:
\begin{align*}
& \E[\mathbf{B}_{t+h}\mIcal_{t,b}]=\widehat{\mathbf{b}}_{t+h}, \\
& \E[\widehat{\mathbf{U}}_{t+h} \mIcal_{t,b} ]=\mathbf{A}\widehat{\mathbf{b}}_{t+h}.
\end{align*}

We now derive the different blocks of the covariance matrix.
The cross-covariance between $\mathbf{B}_{t+h}$ and $\widehat{\mathbf{U}}_{t+h}$ is:
\begin{align*}
\operatorname{Cov(\mathbf{B}_{t+h} ,\widehat{\mathbf{U}}_{t+h} \mIcal_{t,b}})  & =
\operatorname{Cov(\mathbf{B}_{t+h}, \mathbf{A}\mathbf{B}_{t+h} + \boldsymbol{\varepsilon}_{t+h}^u \mIcal_{t,b})} \\
& = \operatorname{Cov(\mathbf{B}_{t+h},
	\mathbf{B}_{t+h} \mIcal_{t,b})} \Smb^T + \operatorname{Cov(\mathbf{B}_{t+h} ,\boldsymbol{\varepsilon}_{t+h}^u \mIcal_{t,b}})  \\
&
= k_h(  \Rb[,1] \Smb^T
+ \boldsymbol{M}_{1}) \in \R^{n \times (m-n)}
\end{align*}
where $k_h>0$ is the multiplicative constant  (Sec.\ref{sec:reconc-h}) that 
yields the variance of the forecasts $h$-steps ahead, given
the covariances of the forecasts 1-step ahead.

The covariance of the upper forecasts is:
\begin{align*}
\operatorname{Cov}(\widehat{\mathbf{U}}_{t+h}\mIcal_{t,b}) & = 
\operatorname{Cov}(\Smb \mathbf{B}_{t+h} +  \boldsymbol{\varepsilon}_{t+h}^u, \Smb \mathbf{B}_{t+h} +  \boldsymbol{\varepsilon}_{t+h}^u \mIcal_{t,b})
\\
&=  k_h\Smb \Rb[,1] \Smb^{T} + k_h \Rmb[,1]  
+ \operatorname{Cov(\mathbf{AB}_{t+h} ,\boldsymbol{\varepsilon}_{t+h}^u)}
+ \operatorname{Cov(\boldsymbol{\varepsilon}_{t+h}^u,\mathbf{AB}_{t+h})}\\
&=  k_h (\Smb \Rb[,1] \Smb^{T} +  \Rmb[,1]  
+ \mathbf{AM}_1
+ \mathbf{M}_1^T \mathbf{A}^T). 
\end{align*}

Hence the joint prior (i.e., before observing $\widehat{\mathbf{U}}_{t+h}$) is:
{\footnotesize
	\begin{align*}
	\begin{pmatrix}
	\mathbf{B}_{t+h} \\
	\widehat{\mathbf{U}}_{t+h}\\
	\end{pmatrix} \mIcal_{t,b} &\sim  N
	\begin{bmatrix}
	\begin{pmatrix}
	\mathbf{\widehat{b}}_{t+h}\\
	\mathbf{A} \mathbf{\widehat{b}}_{t+h}\\
	\end{pmatrix}\!\!,&
	\begin{pmatrix}
	k_h\Rb[,1] & & k_h(\Rb[,1] \Smb^T	 + \mathbf{M}_{1})		\\
	k_h (\Smb \Rb[,1] \ + \mathbf{M}_{1}^T) 
	& &
	k_h (\Smb \Rb[,1] \Smb^{T} +  \Rmb[,1]   + \mathbf{AM}_1 + \mathbf{M}_1^T\mathbf{A}^T)
	\\
	\end{pmatrix}
	\end{bmatrix}.
	\end{align*}
}

Now we receive the forecast $\widehat{\mathbf{u}}_{t+h}$  for the upper time series (recall that $\widehat{\mathbf{U}}_{t+h}$ denotes the random variables while $\widehat{\mathbf{u}}_{t+h}$ denotes observations).
We obtain the posterior distribution for the bottom time series
$ P(\mathbf{B}_{t+h}|
\widehat{\mathbf{u}}_{t+h})$
by applying the standard formulas for the conditional distribution of a MVN distribution \cite[Sec.4.3.1]{murphy2012machine}.
To have a shorter notation, let us define:
\begin{align}
\mathbf{G} & =  
\left(k_h(\Rb[,1] \Smb^T	 + \mathbf{M}_{1})\right)
\left(k_h
(\Smb \Rb[,1] \Smb^{T} + \Rmb[,1]  
+ \mathbf{AM}_1
+ \mathbf{M}_1^T\mathbf{A}^T)\right)^{-1} \nonumber \\
& 
= \left(\Rb[,1] \Smb^T	 + \mathbf{M}_{1}\right)
\left(\Smb \Rb[,1] \Smb^{T} + \Rmb[,1]  
+ \mathbf{AM}_1
+ \mathbf{M}_1^T\mathbf{A}^T\right)^{-1},
\label{eq:G}
\end{align} 
where $k_h$ disappears from the expression of $\mathbf{G}$.

The reconciled bottom time series have then the following mean and variance: 
\begin{align}
& \bT_{t+h}  = \E\left[\mathbf{B}_{t+h}  \mIcal_{t,b}, \uH_{t+h}\right] = \bH _{t+h} + \mathbf{G} (\uH_{t+h} - \Smb\bH_{t+h})
\label{eq:reconciliationKalman}
\\
& \var\left[\mathbf{B}_{t+h} \mIcal_{t,b}, \uH_{t+h} \right] 
=  k_h\left(\Rb[,1] - \mathbf{G} (\Smb \Rb[,1] \ + \mathbf{M}_{1}^T) \right)	\label{eq:reconciliationVar}
\end{align}

Thus the adjustment applied 
to the base forecasts is  proportional to 
($\uH_{t+h}-\Smb\bH_{t+h}$),
i.e. the difference
between the prior mean and the uncertain observation (i.e., the forecasts)  of the upper time series.
The term ($\uH_{t+h}-\Smb\bH_{t+h}$) is called the \textit{incoherence}  of the base forecasts
in \cite{Wickramasuriya.etal2018}. 
The mean of the reconciled bottom time  series  does not depend on $k_h$, while the variance does.

The reconciled point forecast and the covariance
for the entire hierarchy are:
\begin{align}
& \E[\boldsymbol{Y}_{t+h} \mIcal_{t,b}, \uH_{t+h}] = \yT_{t+h}  = \mathbf{S}\bT_{t+h} \\
& \var\left[ \boldsymbol{Y}_{t+h} \mIcal_{t,b}, \widehat{\mathbf{u}}_{t+h} \right]  = \mathbf{S} \var\left[\mathbf{B}_{t+h} \mIcal_{t,b}, \uH_{t+h}\right] \mathbf{S}^T.
\end{align}

\subsection{Related works and optimality of the reconciliation}
Bayes' rule is a well-known tool for information fusion \cite[Sec. 2]{durrant2008multisensor}, and we apply it for the first time for forecast reconciliation.
We will later prove that  the posterior mean of our approach yields the same point predictions of MinT.

Yet, our algorithm additionally provides the predictive distribution for the entire hierarchy;  we thus call it pMinT, where p stands for \textit{probabilistic}.
We also contribute a novel reconciliation approach based on the linear-Gaussian (LG) model, which is obtained by setting  $\mathbf{M}=\mathbf{0}$ in the definition of $\mathbf{G}$. 
Both pMinT and LG are thus probabilistic reconciliation algorithms.

We point out for the first time
a link between the reconciliation problem and the Kalman filter, whose state-update equation can be derived from the linear-Gaussian model \cite{roweis1999unifying}. 
In particular, Equation~\eqref{eq:reconciliationKalman} has the same structure of the state-update of a Kalman filter.
According to the definition of $\mathbf{G}$ of Eq.\eqref{eq:G},
the LG reconciliation 
corresponds to
the standard Kalman filter \cite[Chap.5]{simon2006optimal},
while the pMinT reconciliation corresponds to a generalized Kalman filter 
which assumes correlation between the noise of the state and the noise of the output \cite[Chap.7.1]{simon2006optimal}. 
Thus future research could
explore the literature of the Kalman filter in order to
borrow ideas for the reconciliation problem. 

The optimality of our approach can be informally proven  by considering that it yields the posterior mean of the reconciled forecasts, which is the minimizer of the quadratic loss under the Gaussian assumption \cite[Chap. 5.7]{murphy2012machine}.  Moreover, it has the same equation of the 
state-update step of the
Kalman filter, which provably minimizes the mean squared error of the estimates without any distributional assumption \cite[Chap.5]{simon2006optimal}.
In Sec. \eqref{sec:proof-minT} we will moreover show that our point predictions correspond to those of MinT, which have been  proven to be the minimizer of the mean-squared error.

\subsection{The covariance matrices $\Rb[,1]$ and $\Rmb[,1]$ }
\label{subsec:covariances}

The element $(i,j)$ of $\Rb[,1]$ is the covariance $\operatorname{Cov}(B^i_{t+1}, B^j_{t+1} \mIcal_{t,b})= \operatorname{Cov}(B^i_{t+1}, B^j_{t+1} \mid \mathbf{B}_{1:t} = \mathbf{b}_{1:t})$, where $\mathbf{B}_{1:t} = \mathbf{b}_{1:t}$ denotes a realization $\mathbf{b}_{1:t}$ of $\mathbf{B}_{1:t}$.
Yet we only have one observation of $B^i_{t+1}, B^j_{t+1}$
\textit{conditional} on $\mathcal{I}_{t,b}$, which prevents estimating the covariance.
We can overcome the problem with the following result, which shows that we can approximate $\Rb[,1]$ by computing 
the covariance of the residuals.  

Let us consider the vectors of bottom time series $\mathbf{B}_{1}, \ldots, \mathbf{B}_{t}$ and the conditional expectation $\widehat{\mathbf{B}}_{t+1} = \E[\mathbf{B}_{t+1} \mid \mathbf{B}_{1}, \ldots, \mathbf{B}_{t}] = \E[\mathbf{B}_{t+1} \mid \mathbf{B}_{1:t}]$. Note that $\widehat{\mathbf{B}}_{t+1}$ is a random vector as we have not yet observed $\mathbf{B}_i$, $i=1, \ldots, t$.  In this section we show 
\begin{align}
\E[\operatorname{Cov}(B^i_{t+1}, B^j_{t+1} \mid \mathbf{B}_{1:t}) ] = \operatorname{Cov}(E^i_{t+1},E^j_{t+1}) \qquad i,j=1,\ldots, n,
\label{eq:covEquality}
\end{align}
where $E^i_{t+1} := B^i_{t+1} - \widehat{B}^i_{t+1}$, $i=1, \ldots, n$ denotes
the residual of the model fitted on the $i$-th time series, for the forecast horizon $t+1$.   If we observe $\mathcal{I}_{t,b}$, then we can approximate $\operatorname{Cov}(B^i_{t+1}, B^j_{t+1} \mIcal_{t,b})$ with $\operatorname{Cov}(B^i_{t+1}-\widehat{b}^i_{t+1},B^j_{t+1}-\widehat{b}^j_{t+1})$, the covariance of the residuals of the models fitted on the bottom time series.

Consider now the conditional covariance on the left side of eq.~\eqref{eq:covEquality}, we have

\begin{align*}
\operatorname{Cov}(B^i_{t+1}, B^j_{t+1} \mid \mathbf{B}_{1:t}) &=
\E\left[ \left( B^i_{t+1}- \E[B^i_{t+1} \mid \mathbf{B}_{1:t}] \right) \left(B^j_{t+1} - \E[B^j_{t+1} \mid \mathbf{B}_{1:t}] \right) \mid \mathbf{B}_{1:t} \right] \\
&= \E\left[ \left(B^i_{t+1}- \widehat{B}^i_{t+1} \right) \left(B^j_{t+1} - \widehat{B}^j_{t+1} \right) \mid \mathbf{B}_{1:t} \right]
\end{align*}

By taking the expectation on both sides we obtain 

\begin{align*}
\E[\operatorname{Cov}(B^i_{t+1}, B^j_{t+1} \mid \mathbf{B}_{1:t})] &= \E\left[ \E\left[ \left(B^i_{t+1}- \widehat{B}^i_{t+1} \right) \left(B^j_{t+1} - \widehat{B}^j_{t+1} \right) \mid \mathbf{B}_{1:t} \right]  \right] \\
&=\E\left[ \left(B^i_{t+1}- \widehat{B}^i_{t+1} \right) \left(B^j_{t+1} - \widehat{B}^j_{t+1} \right)  \right] \\
&= \operatorname{Cov}\left(B^i_{t+1}- \widehat{B}^i_{t+1}, B^j_{t+1} - \widehat{B}^j_{t+1}  \right) = \operatorname{Cov}\left( E_{t+1}^i, E_{t+1}^j \right)
\end{align*}

We thus estimate the covariance matrix $\Rb[,1]$ using  the covariance of the residuals of the models fitted on the bottom time series.
\subsubsection{Computation of  $\Rmb[,1]$}
According to Eq.~\eqref{eq:upperNoise},
\begin{align*}
\boldsymbol{\varepsilon}_{t+1}^u =  \Smb \mathbf{B}_{t+1} -
\widehat{\mathbf{U}}_{t+1},
\end{align*}
whose variances and covariances can be readily computed from the residuals.

\section{Reconciliation of a simple hierarchy}
\label{sec:BasicExample}
We now illustrate 
how the base forecasts interact during the reconciliation
of 
a simple hierarchy.
We consider a hierarchy 
constituted by two bottom time series ($B_1$ and $B_2$) and 
an upper time series $U$.
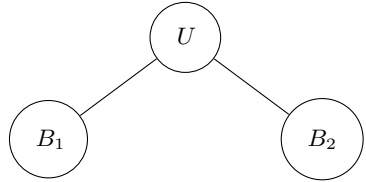
\begin{figure*}[!ht]
	\centering
	\begin{tikzpicture}[level/.style={sibling distance=40mm/#1}, , scale=0.9]
	\node [circle,draw] (z)
	{\textcolor{white}{r}$\, U_{\,}$\textcolor{white}{r}}
	child {node [circle,draw] (a)
		{\textcolor{white}{r}$\,B_1$\textcolor{white}{r}}
	}
	child {node [circle,draw] (j)
		{\textcolor{white}{r}$\,B_2\,$\textcolor{white}{r}}
	};
	\end{tikzpicture}
\end{figure*}

The base forecast for the bottom time series are
the  point forecasts $\widehat{b}_1$ and $\widehat{b}_2$ with variances 
$\sigma_1^2$ and $\sigma_2^2$.
The prior beliefs about $B_1$ and $B_2$ are:
\begin{align*}
\begin{pmatrix}B_1\\
B_2\\
\end{pmatrix} &\sim  N
\begin{bmatrix}
\begin{pmatrix}
\widehat{b}_1\\
\widehat{b}_2\\
\end{pmatrix}\!\!,&
k_h\begin{pmatrix}
\sigma_1^2 & \sigma_{1,2} \\
\sigma_{1,2} & \sigma_2^2,  \\
\end{pmatrix}
\end{bmatrix}
\end{align*}
where  for simplicity we 
remove the forecast horizon ($t+h$) from the notation.

The summing matrix is:
$$\mathbf{S} = 
\begin{bmatrix}
1                 & 1  \\
\cdashline{1-2}
1 & 0   \\
0                 & 1 
\end{bmatrix}
=
\begin{bmatrix}
\multicolumn{2}{c}{\Smb}\\ \cdashline{1-2}
1   & 0   \\
0       & 1 
\end{bmatrix}.
$$

We start considering the simpler case of reconciliation via the LG algorithm. The matrix $\mathbf{G}$ is: 
$$\mathbf{G} =\Rb[,1] \Smb^T (\Rmb[,1] + \Smb\Rb[,1] \Smb^T)^{-1} = \frac{1}{\sigma_u^2 + \sigma^2_1+\sigma^2_2 + 2\sigma_{1,2}} \begin{bmatrix}
\sigma_1^2 + \sigma_{1,2} \\
\sigma_2^2 + \sigma_{1,2},
\end{bmatrix}$$

since  $\Smb=[1\,\, 1]$,  $\Rmb[,1]=\sigma_u^2$ and moreover:
\begin{align*}
& \Rb[,1] \Smb^T = [\sigma_1^2 + \sigma_{1,2} \,\, \, 
\sigma_{1,2} + \sigma_2^2  ]^T,\\
& \Smb^T\Rb[,1] \Smb = \sigma_1^2 + \sigma_2^2 + 2\sigma_{1,2}, \\ 
& \Rmb[,1] + \Smb\Rb[,1] \Smb^T = \sigma_u^2 + \sigma^2_1+\sigma^2_2 + 2\sigma_{1,2}.
\end{align*}
Note that $\mathbf{G}$ does not depend on $h$, as also shown in Eq.~\eqref{eq:G}.

The reconciled bottom forecasts are:
\begin{align}
\bT &= \bH + \mathbf{G}(\widehat{u} - \Smb\bH),
\label{eq:simpleRec}
\end{align}
where $\widehat{u}$ is the base forecast for $U$.

The reconciled bottom forecast can be written as:
\begin{align}
\label{eq:btilde_simple1}
\widetilde{\mathbf{b}} &  = \begin{bmatrix}
\widehat{b}_1 \\ \widehat{b}_2
\end{bmatrix} +  \begin{bmatrix}
\sigma_1^2 +\sigma_{1,2} \\ \sigma_2^2 +\sigma_{1,2}
\end{bmatrix} \frac{\widehat{u} - \Smb^T\mathbf{\widehat{b}} }{\sigma_u^2 + \sigma_1^2 +\sigma_2^2 +2\sigma_{1,2}} \\ \nonumber
\end{align}
Eq.~\eqref{eq:btilde_simple1} shows that the adjustment applied to the base forecasts
depends on $\sigma_u^2$. If $\sigma_u^2$ is 
large the adjustment is small,  since the upper forecast is not informative.
If on the contrary $\sigma_u^2 = 0$, the sum of the reconciled bottom forecasts  is forced to match $\hat{u}$, i.e.,
$\tilde{b}_1 + \tilde{b}_2 = \hat{u}$ (this can be shown by re-working Eq. \eqref{eq:btilde_simple1}).

We now show that the reconciled bottom forecast  are 
a linear combination of the base forecasts. 
Let us define:
\begin{align}
g_1=\frac{\sigma_1^2 +\sigma_{1,2}}{\sigma_u^2 + \sigma_1^2 +\sigma_2^2+2\sigma_{1,2}} 
\,\,\,\,\,
g_2=\frac{ \sigma_2^2 +\sigma_{1,2}}{\sigma_u^2 + \sigma_1^2 +\sigma_2^2+2\sigma_{1,2}} 
\label{g1g2-Bayes}
\end{align}

After some algebra we obtain: 
\begin{align}
\widetilde{\mathbf{b}} &  
=
\begin{bmatrix}
\widehat{b}_1 
\left(
1-\frac{\sigma_1^2 +\sigma_{1,2}}{\sigma_u^2 + \sigma_1^2 +\sigma_2^2+2\sigma_{1,2}} 
\right)
+
(\widehat{u} -\widehat{b}_2)
\frac{ \sigma_1^2 +\sigma_{1,2}}{\sigma_u^2 + \sigma_1^2 +\sigma_2^2+2\sigma_{1,2}} 
\\
\widehat{b}_2 
\left(
1-\frac{\sigma_2^2+\sigma_{1,2}}{\sigma_u^2 + \sigma_1^2 +\sigma_2^2+2\sigma_{1,2}} 
\right)
+
(\widehat{u} -\widehat{b}_1)
\frac{ \sigma_2^2 +\sigma_{1,2}}{\sigma_u^2 + \sigma_1^2 +\sigma_2^2+2\sigma_{1,2}} 
\end{bmatrix} = \begin{bmatrix}
\widehat{b}_1 
\left(
1-g_1 
\right)
+
(\widehat{u} -\widehat{b}_2)g_1
\\
\widehat{b}_2 
\left(
1-g_2 
\right)
+
(\widehat{u} -\widehat{b}_1)g_2, 
\end{bmatrix}
\label{g1g2-LG}
\end{align}

Thus
$\widetilde{b}_1$ is a weighted average of 
two estimates:  $\widehat{b}_1$ and $(\widehat{u}-\widehat{b}_2)$; the weight of $\widehat{b}_1$ decreases with $\sigma_1^2$ and increases with
($\sigma_2^2 + \sigma_u^2)$.

The reconciliation carried out by pMinT is similar to what already discussed, once we 
adopt $g_1^*$ and $g_2^*$ in place of $g_1$ and $g_2$:
\begin{align}
& g_1^*=\frac{\sigma_1^2+\sigma_{1,2} - \sigma_{u,1}}{\sigma_u^2 + \sigma_1^2 +\sigma_2^2+2\sigma_{1,2} -2\sigma_{u,1}-2\sigma_{u,2}} 
\\
& g_2^*=\frac{\sigma_2^2+\sigma_{1,2}-\sigma_{u,2}}{\sigma_u^2 + \sigma_1^2 +\sigma_2^2+2\sigma_{1,2} -2\sigma_{u,1}-2\sigma_{u,2}}  \label{g1g2-Mint}
\end{align}
Thus pMinT accounts also for the cross-covariances $\sigma_{u,1}, \sigma_{u,2}$
between the bottom time series and of noise affecting  the forecasts for the upper time series. 

\subsection{Relationship with MinT}
\label{sec:proof-minT}

Our reconciled bottom time series can be written as: 
\begin{align}
\bT & = \bH + \mathbf{G} (\uH - \Smb\bH) = (\mathbf{I} - \mathbf{G}\Smb)\bH + \mathbf{G}\uH \nonumber \\
& = \begin{bmatrix}
\mathbf{G} & (\mathbf{I}-\mathbf{G}\Smb)
\end{bmatrix}
\begin{bmatrix}
\uH  \\
\bH
\end{bmatrix} = \mathbf{P}_h \yH.
\end{align}
The matrix $\mathbf{P}_h$ of pMinT is thus:
\begin{align}
\mathbf{P}_{pMinT,h}= \begin{bmatrix}
\mathbf{G} & (\mathbf{I}-\mathbf{G}\Smb)
\end{bmatrix}\label{eq:P_pmint}
\end{align}

\begin{proposition}	\label{th:minT-Bayes}
	The matrices $\mathbf{P}_h$ of MinT and pMinT are equivalent.
	The proof is given in the supplementary material.
\end{proposition}

\section{Experiments}
\label{sec:experiments}
In this paper we take a probabilistic point of view; we thus assess 
the reconciled predictive distributions rather than the point forecasts.
Our metric is the \textit{energy score} (ES), a scoring rule for multivariate distributions
\cite{gneiting2007strictly}.
The ES is the multivariate generalization of the 
continuous ranked probability score (CRPS), which is obtained by integrating the Brier score over the predictive distribution of the forecast  \cite{gneiting2007strictly}.
Let $\mathbf{y}$ be the actual multivariate observation, and let us assume that we have
$k$ samples $\mathbf{x}_1, \mathbf{x}_2, \ldots, \mathbf{x}_k$,
from the multivariate predictive distribution $F$.
The energy score is:
\begin{align}
\mathrm{ES}(\mathbf{y},F) = \frac{1}{k}
\sum_{i=1}^{k} \left\lVert  \mathbf{x}_i - \mathbf{y}   \right\rVert
- \frac{1}{2k^2} \sum_{i=1}^{k} \sum_{j=1}^{k}
\left\lVert  \mathbf{x}_i - \mathbf{x}_j   \right\rVert
\end{align}
The energy score is a loss function: the lower, the better.
We consider three methods for probabilistic reconciliation: probabilistic bottom-up (BU), LG and pMinT.
We did not find any package implementing the algorithms of \cite{taieb17a,2014variational}; thus we did not include them in our comparison. 
We estimate all the covariance matrices via the shrinkage estimator \cite{schafer2005shrinkage}.

\paragraph{Base forecasts}
We consider two time series models to  compute the base forecasts.
The first is \textit{ets}, which  fits different exponential smoothing variants and eventually performs model selection via AICc. 
The second method is auto.arima. It first decides how to differentiate the time series 
to make it stationary; 
then, it looks for the best arma model which fits the stationary time series, performing model selection via AICc. Both approaches performed well in forecasting competitions;  they are available from the 
\textit{forecast} package \cite{hyndman2007automatic} for R.
In all simulations, we compute forecasts up to $h$=4.
As no particular pattern exists in the relative performance of the methods as $h$ varies, 
we present the results averaged over $h$=1,2,3,4.

\paragraph{Setting $k_h$}
We are unaware of previous studies on how to set $k_h$.  
In this paper we compare two heuristics, acknowledging that this remains an open problem.
The two options are $k_h$=$h$ and $k_h$=$1$. 
The choice $k_h$=$h$ is based on the following approximation.
The variance of $\widehat{y}_{t+1}$  around $y_{t+1}$ 
is $\sigma^2$; assuming the independence of the errors, 
the variance of $\widehat{y}_{t+1}$  around $y_{t+h}$ is $h \sigma^2$. 
The approximation lies in the fact that we are modeling the variance of $\widehat{y}_{t+1}$ (not $\widehat{y}_{t+h}$)  around $y_{t+h}$. 

Instead, the option $k_h=1$ keeps the variance fixed with $h$. This 
represents the short-term behavior of models
which  contain only seasonal terms and no autoregressive terms.
For instance, when dealing with  a monthly time series, the variance of such models
is constant up to $h$=12.

\paragraph{Code}
The code of 
our experiments is available at: \url{https://github.com/iamthejao/BayesianReconciliation}.

\subsection{Synthetic data}
\begin{figure*}[!ht]
	\centering
	\begin{tikzpicture}[level/.style={sibling distance=40mm/#1}, , scale=0.9]
	\node [circle,draw] (z){Total}
	child {node [circle,draw] (a) {\textcolor{white}{r}$\,$A$\,$\textcolor{white}{r}}
		child {node [circle,draw, fill=gray!20] (b) {AA}
		}
		child {node [circle,draw, fill=gray!20] (g) {AB}
		}
	}
	child {node [circle,draw] (j) {\textcolor{white}{r}$\,$B$\,$\textcolor{white}{r}}
		child {node [circle,draw, fill=gray!20] (k) {BA}
		}
		child {node [circle,draw, fill=gray!20] (l) {BB}
		}
	};
	\end{tikzpicture}
\end{figure*}

We  generate synthetic data sets using the hierarchy above, previously considered in
the experiments of \cite{Wickramasuriya.etal2018}.
We simulate
the four bottom time series as AR(1) processes, drawing their parameters uniformly from the stationary region.
The noises of the bottom time series at each time instant are correlated,  multivariate Gaussian distributed, with mean $\mu = [0,0,0,0]^T$and 
covariance:
$$\Sigma = \begin{bmatrix}
5 & 3 & 2 & 1 \\
3 & 5 & 2 & 1 \\
2& 2 & 5 & 3 \\
1 & 1 & 3 &5 
\end{bmatrix}.$$
Thus $\Sigma$ enforces a stronger correlation between
time series which have the same parents.
At each time instant $t$ we add the noise $\eta_t \sim N(0,10)$ 
to the time series AA and BA  and the noise $(-\eta_t)$ to the time series AB and BB.
In this way we simulate noisy bottom time series ($\eta_t$ and $-\eta_t$ cancel out when dealing with the upper time series) which can be encountered in real cases when several disaggregations are applied to the total time series.
We consider the following length $T$ of the time series: $\{50; 100; 1000\}$.
For each value of $T$ we perform 1000 simulations.
The averaged energy scores are given in Tab.~\ref{tab:synthetic};  in each cell
we report the lower energy score between the case $k_h$=$h$ and $k_h$=$1$.

\begin{table}
	\centering
	\caption{Mean energy score, averaged over 1000 simulations and over $h$=1,2,3,4.  In each row we highlight the lower result.} \label{tab:synthetic}
	\begin{tabular}{rlcrlrrrrr}
		\toprule
		\multicolumn{1}{c}{$T$} &  & \multicolumn{1}{c}{method} &  & $\,\,$ & \multicolumn{1}{c}{$\,\,$BU$\,\,$} &  & \multicolumn{1}{c}{pMinT} &  & \multicolumn{1}{c}{$\,\,$LG$\,\,$} \\ \midrule
		50 &  &           arima            &  &              &                                9.7 &  &                       9.5 &  &                       \textbf{9.4} \\
		50 &  &            ets             &  &              &                                9.9 &  &                       9.8 &  &                       \textbf{9.7} \\
		&  &                            &  &              &                                    &  &  \\
		100 &  &           arima            &  &              &                                9.1 &  &              \textbf{9.0} &  &                       \textbf{9.0} \\
		100 &  &            ets             &  &              &                                9.5 &  &                       9.4 &  &                       \textbf{9.3} \\
		&  &                            &  &              &                                    &  &  \\
		1000 &  &           arima            &  &              &                                8.8 &  &              \textbf{8.7} &  &                                8.8 \\
		1000 &  &            ets             &  &              &                                9.5 &  &              \textbf{9.3} &  &                                9.4 \\ \bottomrule
		&  &                            &  &              &                                    &  &
	\end{tabular}
\end{table}
Since  the time series are stationary, they basically fluctuate around their mean.
In this case the magnitude of the incoherence is generally limited, allowing also the 
bottom-up reconciliation to be competitive.
The ES of both 
pMinT and LG is on average 
1.5\%  smaller than that of BU.  
We also note an advantage of LG over pMinT for small $T$, and instead the reverse
for large $T$; 
this might be the effect of the additional covariances estimated by pMinT (see $\sigma_{u,1}$ and $\sigma_{u,2}$ in Sec.\ref{sec:BasicExample}).

\subsection{Experiments with real data sets}\label{sec:grouped}
We consider two hierarchical time series: \emph{infantgts} and \emph{tourism}. 
Both are \textit{grouped} time series, which is a generalization of hierarchical time series.
In particular 
the time series of a given level are always 
sums of some bottom time series, but they are
not necessarily sums of time series of the adjacent lower level.

\paragraph*{Infantgts}
The \emph{infantgts} is available within the \textbf{hts} \cite{hts} package 
for R.
It contains infant mortality counts in Australia, disaggregated by sex and by eight different states. Each time series contains 71 yearly
observations, covering the period 1933-2003.
The bottom level contains 16 time series (8 states x 2 genders).
The second level  contains 2 time series:  the counts of males and females, aggregated over the states.
The third level sums males and females in each state, 
yielding 8 time series (one for each state). The fourth level is the total. 

\paragraph{Tourism}
The \textit{tourism} data set regards the 
number of nights spent by Australians away from home.
It is available in raw format from 
\url{https://robjhyndman.com/publications/MinT/}.
The time series cover the period 
1998--2016 with monthly frequency.
There are 304 bottom time series, referring 
to 76 regions and 4 purposes. 
The first level 
sums over the purposes, yielding 
76 time series (one for each region); such values are further aggregated 
into macro-zones (27 time series) and states (7 time series). 
Other levels of the hierarchy aggregate the bottom time series of the same zone (yielding 108 time series: 27 zones x 4 purposes), which are then further aggregated into 28 time series (7 states x 4 purposes) and  then 4 time series
(4 purposes). The last  level is the total.
Overall the hierarchy
contains 555 time series.

\begin{table}[!htp]
	\centering
		\caption{Averaged energy scores. Each cell is the average over
		200 reconciliations (50 different training/test splits $\times$ $h$=1,2,3,4) . 
		The rows corresponding to the best-performing values of $k_h$ are 
		highlighted.}
	\label{tab:real-dsets}
	\begin{tabular}{lllclllllrllrllr}
		\toprule
		dset & & & $k_h$     & & & method     & & &       BU & & &    pMinT & & &       LG \\
		\midrule
		infantgts & & & 1 & & &arima & & &  \textbf{334.1} &&&   346.9 & &&   348.5 \\
		& && 1  &&   & ets &&&   334.0 &&&   \textbf{320.0} &&&   334.7 \\
		& & & \cellcolor[HTML]{CBCEFB} $h$ & \cellcolor[HTML]{CBCEFB}&\cellcolor[HTML]{CBCEFB} & \cellcolor[HTML]{CBCEFB} arima & \cellcolor[HTML]{CBCEFB} &\cellcolor[HTML]{CBCEFB} &   \cellcolor[HTML]{CBCEFB}\textbf{327.2} & \cellcolor[HTML]{CBCEFB} & \cellcolor[HTML]{CBCEFB}$\,\,$ &   \cellcolor[HTML]{CBCEFB}335.1 &\cellcolor[HTML]{CBCEFB} $\,\,$ & \cellcolor[HTML]{CBCEFB} &   \cellcolor[HTML]{CBCEFB}331.0 \\
		& & &     \cellcolor[HTML]{CBCEFB} $h$  &\cellcolor[HTML]{CBCEFB} & \cellcolor[HTML]{CBCEFB} & \cellcolor[HTML]{CBCEFB} ets &\cellcolor[HTML]{CBCEFB} & \cellcolor[HTML]{CBCEFB}&   \cellcolor[HTML]{CBCEFB}328.2 & \cellcolor[HTML]{CBCEFB}& \cellcolor[HTML]{CBCEFB}&   \cellcolor[HTML]{CBCEFB}\textbf{313.7} &\cellcolor[HTML]{CBCEFB} & \cellcolor[HTML]{CBCEFB}&   \cellcolor[HTML]{CBCEFB}318.7 \\
		& & & & & & & & & & \\
		tourism & & & \cellcolor[HTML]{CBCEFB} $1$ & \cellcolor[HTML]{CBCEFB}& \cellcolor[HTML]{CBCEFB}& \cellcolor[HTML]{CBCEFB}arima & \cellcolor[HTML]{CBCEFB}& \cellcolor[HTML]{CBCEFB}& \cellcolor[HTML]{CBCEFB}2,737.6 & \cellcolor[HTML]{CBCEFB}&\cellcolor[HTML]{CBCEFB} & \cellcolor[HTML]{CBCEFB}\textbf{2,412.0} &\cellcolor[HTML]{CBCEFB} & \cellcolor[HTML]{CBCEFB}& \cellcolor[HTML]{CBCEFB}2,547.4 \\
		& & &  \cellcolor[HTML]{CBCEFB}  $1$  & \cellcolor[HTML]{CBCEFB}& \cellcolor[HTML]{CBCEFB}& \cellcolor[HTML]{CBCEFB} ets &\cellcolor[HTML]{CBCEFB} & \cellcolor[HTML]{CBCEFB}& \cellcolor[HTML]{CBCEFB}2,496.0 & \cellcolor[HTML]{CBCEFB}& \cellcolor[HTML]{CBCEFB}& \cellcolor[HTML]{CBCEFB}\textbf{2,403.7} & \cellcolor[HTML]{CBCEFB}&\cellcolor[HTML]{CBCEFB} & \cellcolor[HTML]{CBCEFB}2,520.1 \\
		&&& $h$ &&& arima &&& 2,785.3 &&& \textbf{2,380.3} &&& 2,448.2 \\
		&&&    $h$  &&& ets &&& 2,527.1 &&& \textbf{2,353.6} &&& 2,410.3 \\
		\bottomrule& & & & & & & & & & 
	\end{tabular}
\end{table}
%
%
We repeat 50 times the following procedure:
split the time series into training and test, using a different split point; 
compute the base 
forecasts up to $h$=4; reconcile the forecasts.
The reconciliation is independently computed for each $h$.
Each value of Tab.~\ref{tab:real-dsets} is thus the 
average over 200 experiments (50 training/test splits $\times$  $h$=1,2,3,4).
On \emph{infantgts}, all the three reconciliation methods perform better with $k_h=h$, probably because
the variance  of the fitted time series models steadily increases with $h$.  
On the contrary, on  \emph{tourism} all reconciliation algorithms perform better with $k_h$=1;
in this case most models only contain 
the seasonal part.
The rows referring to the best values of $k_h$ are highlighted 
in  Tab.~\ref{tab:real-dsets}.

We call 
\textit{setup} the combination of a data set and a forecasting method, such as
\textless infangts,arima \textgreater. 
The pMinT algorithm yields the lowest energy score in most setups; 
in the next section we check whether the differences between methods are significant.

\paragraph{Statistical analysis}
For each setup we perform a significance tests 
for each pair of algorithms (pMinT vs BU, LG vs BU and pMinT vs LG),  using
the Bayesian signed-rank test \cite{benavoli2017time}, which
returns the posterior probability of a method having lower median 
energy score than another (Tab. \ref{tab:p-values}). 
Such posterior probabilities are numerically equivalent to (1 - \textit{p-value}), where \textit{p-value} is the p-value of the one-sided frequentist signed-rank test. 

In most setups (Tab. \ref{tab:p-values}) high posterior probabilities (implying low p-values) 
support the hypothesis of the pMinT having lower energy score than both BU and LG; moreover they also support
the hypothesis of  LG having lower energy score than BU.

\begin{table}[!htp]
	\centering
		\caption{Posterior probability of  the Bayesian signed rank test.}
	\begin{tabular}{llllrlrrr}
		\toprule
		& & & & & \multicolumn{3}{c}{\textit{Posterior probabilities}} \\
		dset    &   $k_h$ & method      &$\,\,$ &  \multicolumn{1}{c}{pMinT \textless BU} &$\,\,\,\,\,\,\,$ &  \multicolumn{1}{c}{pMinT \textless LG} & & \multicolumn{1}{c}{LG \textless BU}\\
		\midrule
		infantgts & 1 & arima & &        0.24 & &        0.02 && 0.47\\
		&      & ets & &        1.00 & &        0.85 && 1 \\
		& h & arima & &        0.89 & &        0.00 && 1.00 \\
		&      & ets & &        1.00 & &        0.00 && 1 \\
		tourism & 1 & arima & &        1.00 & &        1.00 && 1.00 \\
		&      & ets & &        1.00 & &        1.00 && 0.25 \\
		& h & arima & &        1.00 & &        1.00 && 1.00 \\
		&      & ets & &        1.00 & &        1.00 && 1.00 \\
		\bottomrule& & & 
	\end{tabular} 
	\label{tab:p-values}
\end{table}
\paragraph{Meta-analysis}
We now perform a meta-analysis for
for  each pair of algorithms
across the 
different setups, adopting the Poisson-binomial approach \cite{lacoste2012bayesian,corani2017statistical}.
Consider for instance pMinT vs BU.
We model each setup as a Bernoulli trial,  whose possible outcomes are the victory of pMinT or BU. 
The probability of pMinT winning is taken for each setup from Tab.~\ref{tab:p-values} (the probability of BU winning is just its complement to 1).
We then repeat 10,000 simulations, in which we draw the outcome of each setup according 
to the probabilities of Tab.~\ref{tab:p-values}.

We now report the probability of each method outperforming another method in more than half the setups,
based on out of 10,000 simulations.
Both pMinT and LG wins in more than half the setup with probability 1 against BU.
Moreover, there is
0.85  probability of pMinT winning in more than half of the setups against LG.
We thus recommend pMinT as a general default method for probabilistic reconciliation.

\section{Conclusions}\label{sec:conclusions}
We have derived two algorithms (pMinT and LG) based on Bayes' rule for probabilistic reconciliation.
We have also shown a didactic example which clarifies how base forecast and their variances interact during the reconciliation,
In general  pMinT yields better predictive distributions and thus we recommend it as a default.
The LG method can be anyway an interesting alternative when dealing with small sample sizes.
Future research could borrow ideas from the extensive literature of the 
Kalman filter, based on the link we pointed out between reconciliation and 
Kalman filter

\section*{Acknowledgements}
We acknowledge  support from  grant n.~407540\_167199~/~1 from Swiss NSF (NRP~75 Big Data).

\end{document}